\documentstyle[epsfig]{article}
\newcommand{\pbp}{\bar{p}p}
\newcommand{\pbn}{\bar{p}n}
\newcommand{\NNB}{N\bar{N}}
\newcommand{\NB}{\bar{N}}

\title{Nuclear Neutron Haloes as Seen by Antiprotons}
\author{S. Wycech\thanks{{\it Internet address:} wycech@fuw.edu.pl}
		      and
   R. Smola\'nczuk\thanks{{\it Internet address:} smolan@fuw.edu.pl}\\[2mm]
So{\l}tan Institute for Nuclear Studies,\\
Ho\.za 69, PL-00-681 Warsaw, Poland}
\date{}
\begin{document}

\maketitle
\begin{abstract}
Nuclear interactions of antiprotons in atomic states
are discussed. The total
as well as partial widths for single nucleon capture
events are calculated.
These are compared to the X-ray and recent single
nucleon capture data. The
rates of the neutron or proton captures test nuclear
density distributions at
the extreme nuclear surface. Recently found cases of
neutron and proton haloes
are analysed.
\end{abstract}

\section{ Introduction}

It has been known for a long time that hadronic atoms
are a way to test
the nuclear surface region: the tail of nuclear density
distribution, its
isospin structure and correlations [1]. There
are two methods to
learn these properties:

   1) {\it Measurements of the X ray cascade in hadronic
atoms that provide atomic
levels and widths.} Some fractions of the level energies
are due to nuclear
interactions and some parts of the widths are due to
nuclear captures of the
hadron. For highly excited atomic states the nuclear
effects are small, for
low levels the nuclear capture probability increases
rapidly with decreasing
orbital radii and the cascade terminates suddenly.
These natural limitations
allow to measure one level width per atom. Only in
some special cases two
widths and one level shift may be obtained. The levels
in question are of
large angular momenta that locate the nuclear
interactions on the nuclear
surface. At first, the atomic data are used to learn the
strength and form of
hadron optical potentials. Next, some of the level widths may be used to test
the nuclear density tail.
The shifts are usually difficult to interpret and provide a check on the
optical potential.

   2) {\it Measurements of the nuclear capture products.} A unique detrmination
of
the emitted particles may discriminate captures on protons from captures on
neutrons and signal nuclear correlations. Many experiments using various
detection techniques have presented data, in principle, more informative
than the X-ray data. Unfortunatly, these are also more difficult to interpret
as the initial atomic states are not known and the final state interactions,
in particular the charge exchange reactions, are uncertain.

Some related results in the $\bar{p}$ and other atoms are discussed in
reviews [2,9] and other talks at this meeting. This paper discusses
the neutron density distributions tested by recent antiprotonic CERN
experiments, [3]. The latter are of the
second kind and follow previous antiprotonic [4] and kaonic
[5] studies. Now, however, the separation of $\pbp$ and $\pbn$
annihilation modes is done in a different way. Instead of the final state
mesons it is the final nucleus that is detected by radiochemical methods.
In this way
the rates of reactions $\bar{p}(N,Z) \rightarrow mesons(N-1,Z)$ and
$\bar{p}(N,Z)
\rightarrow mesons(N,Z-1)$ are found. The ratio of these rates allow to detect
the
number of neutrons relative to the number of protons in the region of nuclei
where the
$\bar{p}$ capture occurs. The measurements done in nuclei where the method
is applicable yield results that differ widely. For some nuclei ($\rm{}^{58}
Ni, ^{96}Ru$)
one finds the $n/p$ ratio about unity, in heavy nuclei ($\rm{}^{176} Yb,
^{232}Th, ^{238}U$) it is as
large as 5 or 8 while in $\rm{}^{144} Sm$ it happens to be significantly less
than
unity. Once the final and initial states are known and the reaction mechanism
is understood one can determine where the capture occurs and what nuclear
region is tested by these experiments. Then one can interpret the $n/p$
ratio in terms of "neutron halo" or "neutron skin " and give more precise
meaning to these terms. The purpose of this work is the presentation of the
basic elements of such an analysis [6].

The known difficulties: uncertainty of the capture state and the
necessity to describe the final state interactions are still present. In
particular, the radiochemistry is selective, only cold final nuclei are
seen and a proper description of the final states becomes a question.
Fortunately an additional constraint follows from the measurement itself. It
is given by the ratio of two capture rates: the rate for single nucleon
captures that end with cold residual nuclei and the total capture rate.
The former, cold captures, ammount to 10-20 percent of the total,
and are almost independent on the nucleus. If the capture occurs from a
definite atomic state the total rate is given by the level width and is
in principle measurable by the X-rays. Such clean experiments are not
feasible, yet. The chances of nuclear capture from various initial $\bar{p}$
states have to be calculated with an optical potential derived from other
atomic data. An experimental check is expected with forthcoming CERN
experiments which hopefuly will supply also the transition probabilities per
single stopped antiproton  [6].

Next sections describe briefly: the optical potential, final state mesonic
interactions and the interpretation of the neutron haloes.

\section{ Nuclear Interactions of Atomic Antiprotons}

Antiprotons captured into atomic orbits cascade down, emitting Auger
electrons and X-rays, to be finally absorbed by the nucleus. This happens
in atomic states of high angular momenta, presumably in circular orbits.
Such a scenario is formed by the atomic cascade that tends to populate states
of high angular momenta $l$. Because of the centrifugal barriers and large
${\NNB}$ absorptive cross sections, the nuclear interaction of $\bar{p}$ is
rather well localised at distances as large as twice the nuclear radius.
An important aspect of such peripherality is that it allows low density
approximations in the theoretical description: quasi-free interactions and
a single particle picture of the nucleus. It also facilitates the
description of final mesons, an important issue in understanding of the
absorption experiments. On the other hand, the surface studies are complicated
by the sensitivity of results to an uncertain range of hadron-nucleon forces.

Here, we present a phenomenological description of the antiproton
absorption by nuclei. The level widths are calculated in terms of $\bar{p}$
optical potentials. The simplest one, linear in nuclear density, is
of the form [7]
\begin{equation}
\label{1}
      V^{\rm opt}({\vec R})=\frac{2\pi}{\mu_{\NNB}}t_{\NNB}\rho({\vec R})
\end{equation}
where $\mu_{\NNB}$ is the reduced mass, ${\rho({\vec R})}$ is a nuclear density
and $t_{\NNB}$ is a complex scattering length. The density ${\rho({\vec R})}$
in Eq.(\ref{1}) is not the "bare" nucleon density ${\rho_{0}({\vec R})}$
but a folded one
\begin{equation}
\label{2}
  \rho({\vec R)}=\int d{\vec u}\rho_{0}({\vec R}-{\vec u}) \upsilon({\vec u})
\end{equation}
where $\upsilon$ is a formfactor that describes the $\NNB$ force range. The
length $t_{\NNB}$ in Eq.(\ref{1}) is extracted from the most precise X-ray
measurements done in oxygen isotopes, [7]. The best fit to
these data yields $t_{\NNB}$ of about $-1.5-i2.5$ fm, [7,8].
The corresponding potential is deep and black. For the central densities, $\Im
V^{\rm opt}$
is $200$ MeV strong and the mean free path is well below 1 fm. However,
both the form and the strength of $V^{\rm opt}$ are tested only in the
surface region. Thus, $\Im V^{\rm opt}$ is determined by the atomic level
widths, given by
\begin{equation}
\label{3}
  \Gamma=4\frac{\pi}{\mu_{\NNB}}\Im t_{\NNB}
       \int d{\vec R} \rho({\vec R})\mid\Psi_{\NB}({\vec R})\mid^{2}
\end{equation}
where $\Psi_{\NB}({\vec R})$ is the atomic wave function. Since  $\Psi_{\NB}
\approx{\ R}^l$ is determined essentially by the angular momentum $l$ and
only high momentum states are available the absorption strength is peaked at
the surface.

A typical nuclear interaction region in $\rm{}^{58} Ni$ is shown in Fig.1,
where the
absorption density $W=\rho\mid\Psi\mid^{2}R^{2}$ is plotted. There are two
special atomic states in the capture process. One is called an "upper"
level which usually is the last one that can be detected by the X rays before
the cascading down $\bar{p}$ is absorbed. Its width is obtained from the
intensity loss of X-ray transitions. In this and in many other nuclei the
absorption is most likely to happen from this level. The next circular state
below it, "the lower state", may be reached in some nuclei. In such a case one
measures the shape of X-ray line and extracts the level width and shift.
Nuclear absorptions may happen also in higher atomic orbits in a way that is
not detected by X-ray studies. The chances for the $\bar{p}$ to reach the low
levels in question are not known well.

\begin{figure}[htb]
\epsfxsize=6.0cm \epsfbox{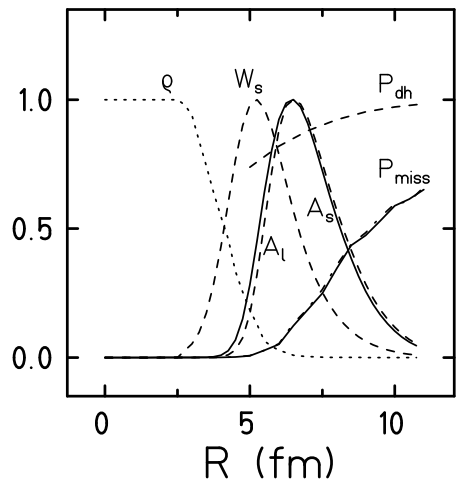}
\caption{
The antiproton absorption densities from n=6, l=5 (upper level)
in $\rm{}^{58} Ni$:  $W_{\rm s}$ for the $\NNB$ annihilation range $r_{\rm
o}=.75$ fm, $\rho$
is a "bare" neutron density.
The cold antiproton absorption densities on a neutron (integrand in
Eq.(\protect\ref{4})): $A_{\rm l}$ for the $\NNB$ annihilation range $r_{\rm
o}=1$ fm and
$A_{\rm s}$ for the range $r_{\rm o}=.75$ fm. Normalisations are arbitrary.
Missing probabilities (left scale): $P_{\rm miss}$ continuous
is due to phase space alone, $P_{\rm miss}$ dash-dotted is calculated
with corrections for the experimental pion momentum distribution.
The flat dashed curve is $P_{\rm dh}$ from the HF model.
}
\end{figure}

The peripherality of capture depends on the range of $\NNB$ forces.
The range parameters in Eq.(\ref{2}) may be adjusted to fit the atomic and low
energy scattering data. Gaussian profile formfactors $\exp(-(r/r_{\rm o})^2)$
have
been used, [8], and typical best fit values are:
$r_{\rm oi}\approx 1$ fm (for $\Im V$) and $r_{\rm or}\approx 1.5$ fm (for $\Re
V$).
On the other hand, calculations based on the ${\NNB}$ potentials yield
values $r_{\rm oi}$ of 0.7 fm up to 1.5 fm, [11], for different
${\NNB}$ states and different ways to go off-shell. The latter values
are probably the upper and lower limits of $r_{\rm oi}$, while the best fit
number is located in-between. An effect of the range uncertainty is shown in
Fig.1 for a partial decay width. The effect on the full width is essentially
the same.

Optical model calculations based on the ${\NNB}$ interaction potentials
[10,11] show that the lengths $t_{\NNB}$ are not the
${\NNB}$ S-wave
scattering lengths. The latter are smaller and repulsive (positive ). The
effective $\Re t_{\NNB}$ is density dependent and has a complicated structure.
At the nuclear surface it reflects a long attractive tail of the pion exchange
forces, at distances around the nuclear radius it may turn to repulsion due
to repulsive scattering lengths, and it is rather uncertain at the nuclear
matter
densities. On the other hand, the phenomenological best fit $\Im t_{\NNB}$
represents cumulative effect of the S and P wave absorptive amplitudes and
can be well understood in terms of the free $\Im t_{\NNB}$. Theoretical
optical potentials are more complicated than formula (\ref{1}), but do not
reproduce the data as accurately as the latter with the best fit parameters.

The level widths reflect all modes of nuclear absorption. The initial
stage of this process, an elementary ${\NNB}$ annihilation, generates
$\approx 2$ GeV energy that
turns mostly into kinetic energy of the final state mesons. These mesons excite
residual nuclei by inelastic processes. To calculate the widths one sums over
unobserved nuclear excited states. The large energy release allows a closure
over these states and the effective $\Im t^{\NNB}$ is close to the
absorptive part of the free $\Im t^{\NNB}$. This is not true when
final nuclear states are limited by the measurements. Thus, the
radiochemical detection allows only cold nuclei i.e. nuclei either in the
ground states or excited less than the neutron separation threshold,
[3].
The effect of this limitation is discussed in the next section.

\section{ Nuclear ${\NNB}$ Annihilation and Final State Interactions}

The aim of this section is to calculate the rate of nuclear $\bar{p}$
annihilations that lead to cold final nuclei. This is done in few steps:

1) An amplitude for the ${\NNB}$ annihilation into mesons
$t_{\NNB \rightarrow M}$ is assumed and introduced into the nuclear
transition amplitude in the impulse approximation.

2) The emission probabilities are calculated and summed over mesonic
and nuclear final states. For an isolated ${\NNB}$ annihilation this
procedure would produce the absorptive cross section and via unitarity
condition the absorptive amplitude $\Im t_{\NNB}$. For nuclear captures
leading to cold nuclei we limit the summation over final states to the
states of elastic meson nucleus scattering. This limited summation generates
the $\Im t_{\NNB}$ again, but now it is folded over nuclear final state
interaction factors, $P_{\rm miss}({\vec R})$, which describe the probability
that the annihilation mesons born at point ${\vec R}$ miss the residual
nucleus.

3) Let a nucleon $N$ occupy a single particle level $\alpha$ with
a wave function $\varphi_{\alpha}({\vec X})$. Then, the ${\bar{p}N}$
annihilation process limited by $P_{\rm miss}$ leads to a single hole final
nuclear state. The experimental condition allows only those initial levels
$\alpha$ that end in cold final nuclei. A factor
\mbox{$P_{\rm dh}(\vec X)=\sum_{\alpha}^{ltd} \varphi_{\alpha}^2/ \sum_{\alpha}
\varphi_{\alpha}^2$ } accounts for this condition.

 4) The finite range effects are described by separable potentials as
done in [11].

The result for a partial width corresponding to a cold capture on a nucleon
$s=$ (proton or neutron) is now
\begin{equation}
\label{4}
  \Gamma_{s}(cold)=4\frac{\pi}{\mu_{\NNB}}\Im t^s_{\NNB}\\
       \int\mid\Psi_{\NB}({\vec Y})\mid^{2}
       \upsilon({\vec Y}-{\vec X})
       \rho_{s}({\vec X})P_{\rm dh}({\vec X})P_{\rm miss}({\vec R})
\end{equation}
where $ t^s_{\NNB} $ is an effective length and ${\vec R}$ is the birthplace
of the mesons. The assumption, justified later, is that all the mesons are
emitted from the central point of the annihilation region
${\vec R}=\frac{{\vec X}+{\vec Y}}{2}$.

Eq.(\ref{4}) is the result, now we explain briefly the final state interactions
that determine $P_{\rm miss}$ and in the next section we turn to calculations
of the nuclear densities.

The spectrum of mesons consists essentially of pions correlated in a sizable
fraction into $\rho$ and $\omega$. These heavy mesons propagate some
\mbox{1 fm} and then turn into pions. The pion multiplicities range
from 2 to 8 with an average 4$-$5 and an average momentum as large as 400 MeV.
Nuclear interactions of these pions may be absorptive, inelastic or elastic.
All the absorptive and almost all the inelastic processes would not leave
the
residual nuclei in cold states and so the production rate for the latter is
given essentially by the elastic scattering. This allows an optical
potential description. In addition, in the bulk of phase space the pions are
fast enough to allow also an eikonal description. Following this, the wave
function for each pion is taken in the form
\begin{equation}
\label{5}
      \bar{\psi}^{(-)}({\vec p}{\vec \xi}) =
      \exp(i{\vec p}{\vec \xi}-iS({\vec p},{\vec \xi}))
\end{equation}
where ${\vec p}$ is a momentum, ${\vec \xi}$ is a coordinate and $S$ is
expressed in
terms of the pion-nucleus optical potential
\begin{equation}
\label{6}
	 S({\vec p},{\vec \xi}) =
  \int_0^{\infty} ds (\sqrt{(p^2-U^{\rm opt}({\vec \xi}+\hat{{\vec p}}s)}-p)
\end{equation}
The $S$ is calculated by integrating the local momentum over the stright
line trajectory. Due to nuclear excitations
and pion absorptions this wave is damped with a rate described by $\Im S$.
The latter is generated by absorptive part of the pionic optical potential
$\Im U^{\rm opt}$. This damping follows the whole path but the main
effect comes from regions of large nuclear densities and not the region
around the birth place ${\vec \xi}$. We assume that all functions
$S({\vec p},{\vec \xi})$ are related to the central point of annihilation
${\vec R}$ that is the $\NNB$ CM coordinate. With the mesonic wave
functions (\ref{5}) the dependence on the total momentum of mesons ${\vec P}$
factorizes
approximately to a plane wave form. One consequence is that the ${\NNB}$
CM "conservation" $\delta({\vec R}-{\vec R'})$  arises in the transition
probabilities
integrated over final momenta. Now, the final state
pion interaction factors may be collected into a probability distribution
\begin{equation}
\label{7}
  P_{\rm miss}({\vec R})=< \prod\mid\exp(-S({\vec p_i},{\vec R}))\mid^2 >
\end{equation}
It is a product of the eikonal factors  averaged over the number and
phase space of final pions with some allowance for an unknown momentum
dependence generated by $t_{\NNB \rightarrow M}$.

The calculations of $P_{\rm miss}$ are performed in a Monte Carlo procedure.
The optical potential for pions must cover wide momentum range from the
threshold up to 0.9 GeV but the phase space favours a region just above
$\Delta$ resonance. This potential is related to the pion nucleon forward
scattering amplitudes and in this way to the pion nucleon cross sections.
That method is established around the $\Delta$ [13].
Here, it is extended to higher $N^{*}_{11}$, $N^{*}_{13}$ resonances which
are described by Breit-Wigner
amplitudes. The two nucleon absorption mode is taken in a phenomenological
form [14]. Performing these calculations one finds that:
high energy expansion of the square root in Eq.(\ref{6}) is satisfactory,
higher resonances cannot be neglected and the black sphere limit is a good
approximation in dense regions. The result is close to a pure geometrical
estimate that relates $P_{\rm miss}({\vec R})$
to the solid angle of the nucleus viewed from the point ${\vec R}$ [9].
In the surface region of interest the $P_{\rm miss}$ is a linear function of
the
radius. It is very fortunate, it makes this calculation
fairly independent on the size of the $\NNB$ annihilation region and on the
uncertain range of the heavy meson propagation.

Examples of $P_{\rm miss},P_{\rm dh}$ and the density $A$ for cold absorption
generated by Eq.(\ref{4}) are given in Fig.1. The latter is seen to be
more peripheral than the total absorption density $W$. The uncertainty
due to the $\NNB$ force range is rather small, beeing moderated by effects of
the strong absorption. The uncertainty of the initial atomic state is shown
in Table 1. The ratio of cold to total capture rates
$\Gamma^{\rm c}/\Gamma^{\rm t}$ raises quickly with the increasing angular
momentum.
The experimental results, given in Table 2, indicate that a significant
contribution of high $l$ states is unlikely. On the other hand, some
participation of lower $l$ states is possible. However, due to the atomic
cascade properties, that seems to be unlikely. A definitive answer requires
experimental and theoretical studies which are beeing undertaken, [6].

\begin{table}[htb]
\caption{Column 2 contains main quantum number n for the lower and upper
circular states.
The weigths give probabilities for nuclear capture calculated under the
assumption that the circular atomic state ${\rm n}={\rm n}_{\rm upper}+1$ is
fully occupied.
The absorption widths: total $\Gamma^{\rm t}$ and cold $\Gamma^{\rm c}$ are
calculated
with the AD model, $R_{np}=.63$. }

\begin{center}
\begin{tabular}{||r|r|r||r|r||}
\hline
ELEMENT & n & weight & $\Gamma^{\rm c}/\Gamma^{\rm t}$ &
$\Gamma_{n}/\Gamma_{p}$ \\
\hline
               	&    &     &      &        \\
 $\rm{}^{58} Ni$&  4 &  0  & .095 &  .69   \\
		&  5 & .16 & .097 &  .69   \\
		&  6 & .83 & .11  &  .70   \\
		&  7 & .01 & .15  &  .71   \\
		&  8 &  0  & .22  &  .71   \\
 $\rm{}^{90} Zr$&  6 & .24 & .106 & 4.67   \\
		&  7 & .72 & .128 & 5.30   \\
 $\rm{}^{154} Sm$&  7 & .01 & .087 & 3.65   \\
		&  8 & .75 & .099 & 3.98   \\
 $\rm{}^{238} U$&  9 & .29 & .106 & 6.55   \\
		& 10 & .71 & .138 & 8.24   \\
		&    &     &      &        \\
\hline
\end{tabular}
\end{center}
\end{table}

The absorbtion widths are given by superpositions of high moments of the
nuclear density distributions. We find the $2l-2$ moment as the dominant
one for the total widths and the $2l$ moment to dominate the cold capture
width. The "neutron halo " is thus understood as a ratio of these high moments
of neutron and proton density distributions. The "neutron skin " is related
the m.sq. radius or other low moments.

\section{ Nuclear Densities}

     As the simplest estimate we use an asymptotic density (AD) model. It
follows the Bethe-Siemens approach [12] although the larger input includes:
charge density distribution, neutron and proton separation energies
and difference of the  rms  radii of proton and neutron density distributions.
At central densities  a  Fermi  gas  of protons and neutrons  is  assumed.
The  Fermi  momenta  are
determined by the densities and Fermi energies are fixed by  the  separation
energies. This gives depth of the potential well which  in  the  surface
region is extrapolated down in the Woods-Saxon form. The  densities  are
given by the exponential damping of the nucleon wave  functions  due  to
the potential barriers. For protons  a  Coulomb  barrier  is  added  and
potential parameters $(c,t)$ are fitted to reproduce the experimental charge
density down to 5 percent of the central density. For neutrons the  same
$t$ is used but $c$ is chosen to obtain the rms radius equal (or larger
by 0.05 fm in the heaviest nuclei) to the proton density rms radius. This
model is expected to generate average level densities, it misses shell
effects and correlations.

     A second model used to determine neutron and proton densities
is the Hartree-Fock (HF) and the Hartree-Fock-Bogolyubov (HFB) scheme with
the effective two-body Skyrme-type interaction. Our aim in using HF
and HFB methods to find nucleon densities at the
extreme tails of the nuclear matter distribution (at distances of 8$-$15
fm from the center) is rather unusual. The necessary practical condition is
the use of a HF code not restricting the asymptotic form of s.p. wave
functions. This excludes e.g. all codes using the harmonic oscillator basis.
In the present work we have applied the code solving HF equations on the
spatial mesh, in which all fields and densities are expressed in the
coordinate representation.

     The HF method disregards completely pairing correlations. To
account for them we used the HFB theory [16,17] which unifies the
self-consistent description of nuclear orbitals, characteristic of HF
method, and the mean field treatment of residual pairing interaction
into a single variational theory.
     The effective force is the ten-parameter Skyrme SkP interaction
described in ref [18].  It has a virtue that the pairing matrix
elements are determined by the force itself, contrary to other
Skyrme-type interactions which define only the particle-hole channel.
 The paired HFB ground-state has not the BCS form so there is no
 simple pairing gap parameter although a kind of average gap can be
 defined as the pairing potential average over occupied states.

     The most severe restriction of the presented results is the imposed
spherical symmetry, both in the HF and HFB codes.  It allows enormous
simplification of solutions, in particular the HFB equation takes the
form of two coupled differential equations in the radial variable for
each value of s.p. angular momentum.

     The density matrix is obtained by summing contributions from the
lowest s.p. orbits. The degeneracy of the spherical subshells is handled
by taking contribution of the last orbit in the filling approximation, i.e.,
an appropriate occupation probability smaller than one is, if
necessary, associated with this orbit.

Some results are collected in Table 2. To find $\Gamma_{n}/\Gamma_{p}$
a ratio of $\pbn$ and $\pbp$ absorptive amplitudes $R_{np}$ is needed.
One number $R_{np}=.63$ follows from mesonic studies in $\rm{}^{12} C$
[4]. It depends on uncertain final charge exchange processes and
we use the deuteron value $R_{np}=.82$ [15]. The latter coincides
with a good fit to $\Gamma_{n}/\Gamma_{p}$ in $\rm{}^{58} Ni$.

\begin{table}[htb]
\caption{Experimental and calculated results for $\Gamma^{\rm c}/\Gamma^{\rm
t}$
and $\Gamma_{n}/\Gamma_{p}$. Calculations are averaged over few atomic
orbitals weighted as in Table 1, with $R_{np}=.82$.}

\begin{center}
\begin{tabular}{||r||r|r||r|r||r|r||}
\hline
 & $\Gamma^{\rm c}/\Gamma^{\rm t}$ & $\Gamma_{n}/\Gamma_{p}$ &
  $\Gamma^{\rm c}/\Gamma^{\rm t}$ & $\Gamma_{n}/\Gamma_{p}$ &
  $\Gamma^{\rm c}/\Gamma^{\rm t}$ & $\Gamma_{n}/\Gamma_{p}$ \\
\hline
 & Exp. [3]     &  & Asympt. &  & HF &  \\
\hline
		&          &           &     &      &      &       \\
  $^{58}$Ni     & .098(8)  &  0.9(1)   & .11 & .90  &.110  & .785  \\
  $^{90}$Zr     & .161(22) &  2.6(3)   & .12 & 4.9  &.125  & 2.54  \\
  $^{96}$Ru     & .113(17) &  0.8(3)   & .10 & 1.7  &.099  & .944  \\
 $^{130}$Te     & .184(36) &  4.1(1)   & .12 & 2.6  &.124  & 3.14  \\
 $^{144}$Sm     & .117(20) &$\leq .4\ \ $ & .09 & 1.9  &.094  & 1.38  \\
 $^{154}$Sm     & .121(20) &  2.0(3)   & .10 & 5.1  &.110  & 3.34  \\
 $^{176}$Yb     & .241(40) &  8.1(7)  & .12 & 4.8  &.111  & 3.23  \\
 $^{232}$Th     & .095(14) &  5.4(8)   & .12 & 7.6  &.087  & 3.80  \\
 $^{238}$U      & .114(9)  &  6.0(8)   & .13 & 10   &.092  & 4.09  \\
		&          &           &     &      &      &       \\
\hline
\end{tabular}
\end{center}
\end{table}
\vspace*{0.5cm}
Our conclusions are:

1) The AD model overestimates the neutron haloes. The latter are not
given by the binding energies and Coulomb barriers alone.
The shell effects (angular momentum) are essential, correlations
of HFB type have rather small effect. These conclusions are similar to
the results of subcoulomb neutron pickup studies [19].

2) Even at these nuclear peripheries at least 2-3 nucleon orbitals
are involved in the capture.

3) There are two anomalous cases: Yb and Te. The anomalies are apparently
related to a strong E2 mixing in those atoms.

4) An interesting case of proton halo is seen in $\rm{}^{144} Sm$. It is not
understood yet.

5) The antiprotonic study of nuclear surface is a promising method, despite
some uncertainties.

The authors acknowledge support by KBN Grants 2 P302 010 07 and Pb2 0956 91 01

\end{document}